# PSTN: Periodic Spatial-temporal Deep Neural Network for Traffic Condition Prediction

Tiange Wang, Zijun Zhang, *Senior Member, IEEE*, and Kwok-Leung Tsui

*Abstract*—Accurate forecasting of traffic conditions is critical for improving safety, stability, and efficiency of a city transportation system. In reality, it is challenging to produce accurate traffic forecasts due to the complex and dynamic spatiotemporal correlations. Most existing works only consider partial characteristics and features of traffic data, and result in unsatisfactory performances on modeling and forecasting. In this paper, we propose a periodic spatial-temporal deep neural network (PSTN) with three pivotal modules to improve the forecasting performance of traffic conditions through a novel integration of three types of information. First, the historical traffic information is folded and fed into a module consisting of a graph convolutional network and a temporal convolutional network. Second, the recent traffic information together with the historical output passes through the second module consisting of a graph convolutional network and a gated recurrent unit framework. Finally, a multi-layer perceptron is applied to process the auxiliary road attributes and output the final predictions. Experimental results on two publicly accessible real-world urban traffic data sets show that the proposed PSTN outperforms the state-of-the-art benchmarks by significant margins for short-term traffic conditions forecasting.

*Index Terms*—deep learning, data-driven model, periodic traffic data, spatial-temporal, traffic condition prediction

## I. INTRODUCTION

WITH the acceleration of urbanization and the growth of population density in metropolitan area, a significant amount of attention has been paid to developing intelligent technologies to improve the transportation efficiency [1]. Reliable traffic prediction is critical to develop the intelligent transportation systems (ITS) in cities. Knowing the traffic condition in advance can assist individuals with effective route planning as well as help authorities to re-route vehicles and mitigate traffic congestion [2].

Most of the existing work on the traffic prediction focus on the evaluation of parameters related to traffic conditions in the short-term future. For example, predicting the traffic flow or volume at a fixed location [3, 4] or predicting the traffic speed of future periods on a targeted road segment [5]. Although these parameters are valuable to monitoring traffic conditions, identifying the congestion level in the next few minutes or hours for road segments provides a more straightforward description of the traffic condition. In this paper, we thus aim to directly predict traffic congestion levels in urban road networks. The prediction of traffic conditions is a challenging problem affected by following factors:

- Spatial dependency. The traffic condition at target road segment is spatially correlated with nearby locations. Considering the complexity of road crossings or lanes, the interplay between roads is hard to discern. For example, the traffic congestion caused by incidents on one road crossing may have impacts on not only neighboring places but also distant roads in a future period, leading to dynamic local coherence in space. Fig. 1 shows the traffic conditions of road segment $v_i$ and other road segments.
- Temporal dependency. The traffic congestion levels at the adjacent time intervals have a strong correlation with each other, and the correlation diminishes as the temporal distance increases. Moreover, in long-term patterns, traffic data usually presents periodicity associated with the closeness, period and trend. For example, traffic conditions during the morning rush hours are similar on consecutive workdays but are different from the same time slots on weekends or other time slots of the same day.
- Geographical factors. Traffic condition is also influenced by the geographical structure of roads. Road attributes also determine the congestion level to a certain extent. For example, the traffic on a main road with more width and length would be different from that of a lane; the speed limit of the roads also has an impact on the traffic flows and furthermore affects the traffic conditions.

These characteristics imply that the correlations both in space and time are not globally the same, which remains the prediction of traffic condition increasingly challenging. Over the last few decades, many attempts have been done to predict traffic condition with only the use of time series [6] or just a little of spatial information [7], but very few of them consider

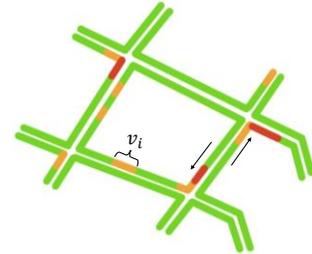

Fig. 1. An example of traffic conditions on road segment $v_i$ and other road segments.

T. Wang and Z. Zhang are with the School of Data Science, City University of Hong Kong, Hong Kong. (Email: zijzhang@cityu.edu.hk)

K.-L Tsui is with Grado Department of Industrial and Systems Engineering, Virginia Polytechnic Institute and State University, Blacksburg, VA, USA.

the similar patterns on historical weekly periods and recent time slots at the same time. External factors, such as attributes of road segments, are ignored as well in modeling the traffic condition, which may degrade the prediction accuracy. To address the above problems, we propose a periodic spatial-temporal deep neural network (PSTN) for the traffic condition prediction task based on the urban road network and periodic traffic data. It integrates graph neural layers, sequence neural layers, and feedforward layers to model the complex relations between road segments from both spatial and temporal aspects. The main contributions of this paper are as follows:

1. Different from previous methods that model only local temporal dependency of traffic patterns, we propose to separate the temporal dependency to a two-scale input, which considers traffic data from historical weekly periods and recent time sequence simultaneously. By aligning spatial information with the two-scale temporal features, we are thus able to predict future congestion levels different steps ahead.
2. Besides the combination of spatial-temporal information, road attributes, such as the length and width, are also concatenated with the output from last module to enhance the prediction of the specific road segment. Via a multi-layer perceptron (MLP) followed the concatenation, the method can predict the congestion level directly.
3. Computational experiments are conducted using two different datasets of records in Beijing and Xi'an cities in Mainland China, respectively. The results show that the proposed PSTN owns a remarkable improvement compared with a number of benchmarks. Additionally, ablation studies exploring the effectiveness of different modules is organized to demonstrate the necessity of aligning spatial and temporal dependencies.

## II. RELATED WORK

Traffic condition prediction is a pivotal application in ITS. It plays an important role in urban traffic control and smart city development. The methods of traffic prediction have undergone different stages of evolution. Traditional approaches for traffic predictions can be divided into classical statistical models, which rely on the time-series analysis of the data regularity, and machine learning methods, which develop nonparametric models to capture the non-stationary characteristics of traffic data.

In statistical models, the autoregressive integrated moving average (ARIMA) [8] was one of the most widely applied fundamental works for traffic predictions. Parameterized by autoregressive terms, non-seasonal differences, and lagged forecast errors, ARIMA was adopted to regress previous values and furthermore realize the short-term traffic prediction. After that, the variants of ARIMA model including Kohonen ARIMA [9], space-time ARIMA [10] and seasonal ARIMA [11] were subsequently developed to improve the prediction performance. However, the statistical models are pre-determined based on an ideal stationary assumption that the traffic data is small and less dynamic, while in the real world, there exist great nonlinear and uncertain characteristics in traffic data. In addition, since the statistical models only consider the temporal information, the spatial dependency of traffic data is ignored.

Machine learning methods emerged to model more complicated characteristics in traffic data. In [12], a regression model based on human-engineered features was trained to predict the traffic demand. In [13], a hybrid model was constructed based on the Gaussian process and tracking algorithm to accurately predict the traffic multi-step ahead. In [14], a multi-step traffic condition prediction model based on the K-nearest neighbors (K-NN) algorithm with the utilization of GPS data from taxis was also developed. In [15], the support vector machines (SVM) regression theory and Chaos-Wavelet Analysis was integrated to capture the non-stationary characteristics of traffic speed data with a new kernel function. Similarly, the support vector regression model (SVR) and particle swarm optimization (PSO) were combined in [16] as a hybrid PSO-SVR forecasting method for the traffic flow. The results showed that the PSO-SVR method performed robustly based on the traffic data containing noises. Even though machine learning methods can provide predictions in some practical situations, their successes heavily depend on the features engineered with strong domain knowledge and model structures developed by them might not be flexible enough to capture nonlinear patterns. Thus, more advanced data-driven methods are required for modeling complex and dynamic traffic data.

In recent years, deep learning models with remarkable capability of exploiting complex and nonlinear patterns have attracted the attention of researchers to apply them on traffic prediction. Some of deep learning methods consider the temporal dependency only. For example, in [17], a deep architecture consisting of a deep belief network (DBN) and a regression layer was proposed to capture random features from multiple traffic datasets. Recurrent neural network (RNN) is a classical type of deep neural networks which is designed for modeling the temporal information. However, the vanilla RNN suffers from the vanishing gradient problem during the backpropagation, which makes RNNs only have the short-term memory. To address this problem, the variants of RNN including the long short-term memory (LSTM) [18-20] and the gated recurrent unit (GRU) [21] were applied with the self-circulation mechanism to better extract the temporal dependency of traffic data. In [22], a capsules network and temporal convolutional network (TCN) was employed as the basic unit to learn the spatial dependence, time dependence, and external factors of the traffic flow prediction. These methods treat traffic sequence data from different road segments independently and are not sufficiently utilizing the spatial information from the urban road network. Thus, convolutional neural networks (CNNs) were introduced to extract the spatial information and combined with sequence models [6, 7, 23], which improved the traffic prediction accuracy. In [24], a stacked autoencoder (SAE) model was applied to capture spatial and temporal correlations and realize the short-term traffic flow prediction. In [25], a deep learning method called the fusion convolutional long short-term memory network (FCL-Net) was developed. The spatial dependency, temporal dependency, and exogenous dependency were taken into account for the short-term passenger demand forecasting. Considering that CNN is limited to modeling the Euclidean space and cannot essentially characterize the spatial dependency with the non-Euclidean topology, graph convolutional networks (GCNs) emerged to capture the spatial information on traffic networks [5]. Representative models such as the diffusion convolutional recurrent neural network (DCRNN) [26], temporal graph convolutional network (T-GCN) [27], and spatial–temporal 3D convolutional neural network (ST-3DNet) [4] have been proposed to capture both the spatial and temporal information. They typically combine the variants of RNN and GCNs to model the spatial-temporal relations in traffic data.

Although existing spatial-temporal models have remarkably improved the performance on the traffic prediction, they still suffer from following limitations. One is that the relations between the roads in the network are much more complicated and dependent on various factors, such as the speed limit, number of lanes, width and length, etc. Meanwhile, the RNN based temporal processing is required to focus on recent periods and is harder to learn the temporal dependency effectively as the sequence go longer. To address the aforementioned challenges, in this paper, we propose a novel spatial-temporal network to directly predict the traffic condition in terms of the congestion level. Besides the regular spatial and temporal dependencies, the proposed method also utilizes historical weekly periods and road attributes as inputs in the model development.

## III. METHOD DESCRIPTION

### A. Problem Definition

In the problem of the traffic condition prediction, we intend to directly predict the congestion level in a future period based on the traffic information on the roads. Specifically, the traffic information can be represented as three types, the traffic features, road network, and road attributes. The traffic features can be represented as a sequence $X_i: \{x_{i,1}, x_{i,2}, ..., x_{i,\tau}\}$, where each element $x_{i,t} \in \mathbb{R}^{1 \times d}$ denotes the $d$-dimenional condition-related features of a certain road segment $v_i$ at time $t$. The traffic network is denoted as $G_i: \{V_i, E_i\}$, which is a subgraph centered on road segment $v_i$. $V_i: \{v_1, v_2, ..., v_i, ..., v_N\}$ is a set of $N$ road segments centered on $v_i$ and $E_i$ is a set of edges connecting them. Based on $V_i$ and $E_i$, the subgraph can also be described by a symmetric adjacency matrix $A_i \in \{0,1\}^{N \times N}$ and a feature matrix $\mathcal{X}_i \in \mathbb{R}^{N \times d}$. Each element in $A_i$ represents the geographical proximity of the two segments. Noted that the element value is 1 if the two segments are adjacent, and 0 if they are not adjacent. Each element in $\mathcal{X}_i$ represents the feature vector of one road segment in $V_i$. Finally, the attributes vector of a certain road segment $v_i$ is represented as $AT_i$. The definition of the traffic congestion level is simple. A categorical parameter $y_{i,t}$ is utilized to present the congestion level of road segment $v_i$ at time $t$.

Based on the formerly presented notations, the problem is considered as learning a function $f$ on the basis of $X_i$, $G_i$, and $AT_i$ to predict the traffic condition in the future $T$ time steps. It can be formulated as:

$$[y_{i,1}, y_{i,2}, ..., y_{i,T}] = f(G_i, X_i, AT_i), \quad (1)$$

### B. Methodology

The framework of the PSTN is shown in Fig. 2 (a). It is mainly composed of three parts organized sequentially: 1) the GCN for capturing topological structure of the road network to obtain the spatial relation; 2) the TCN and GRU for capturing the periodic temporal dependency and local temporal dependency, respectively; 3) the MLP for yielding the final prediction with the combination of road attributes. Note that In Fig. 2 (b), the historical input sequence is folded as weekly periods (i.e. the same time from one week ago, two weeks ago, three weeks ago, and four weeks ago are folded together), which are further used to capture the periodic temporal dependency. Moreover, we adopt the historical weekly periods of future time to better reflect the periodicity of traffic data. The length of the historical time ($t$) and the length of the future time ($T$) do not need to be the same. We require $t \geq T$ in the experiments.

In this section, the details of each module will be introduced.

#### 1) Spatial Dependency Modeling

As illustrated in Fig. 2, the periodic traffic sequence and recent traffic sequence are both fed into GCN to model the spatial dependency. Since the traffic condition at a specific road segment can be influenced by the nearby road segments, the local interactions between them can be constructed as a subgraph. Fig. 3 displays the workflow of GCN. Assume that road segment $v_i$ is the central node, the GCN encodes the topological structure of $G_i$ and obtains the spatial dependency at a specific time. To capture the local dependency of topological structures, GCN takes the adjacency matrix $A_i$ and the feature matrix $\mathcal{X}_i$ as inputs. The GCN model in Fig. 3 is developed by stacking two convolutional layers. The gathering information process in each layer in GCN is formulated as follows:

$$\mathcal{X}_l = \sigma_l\left(\widetilde{D}^{-\frac{1}{2}} \widetilde{A} \widetilde{D}^{-\frac{1}{2}} \mathcal{X}_{l-1} W_l\right), \quad (2)$$

where $\mathcal{X}_{l-1}$ represents the output from the last convolutional layer $l-1$, $W_l$ represents the weight matrix of the next layer $l$. Note that $\widetilde{A} = A + I_N$ and $\widetilde{D} = \sum_j \widetilde{A}_{i,j}$ where $A$ is the adjacency matrix and $\widetilde{D}$ is the degree matrix of $A$. To avoid missing the features of $v_i$ itself, an identity matrix $I_N$ is added to get a new adjacency matrix $\widetilde{A}$. Finally, $\sigma_l$ represents the activation function of the convolutional layer $l$. In our experiments, the two layers are both followed by a ReLU activation.

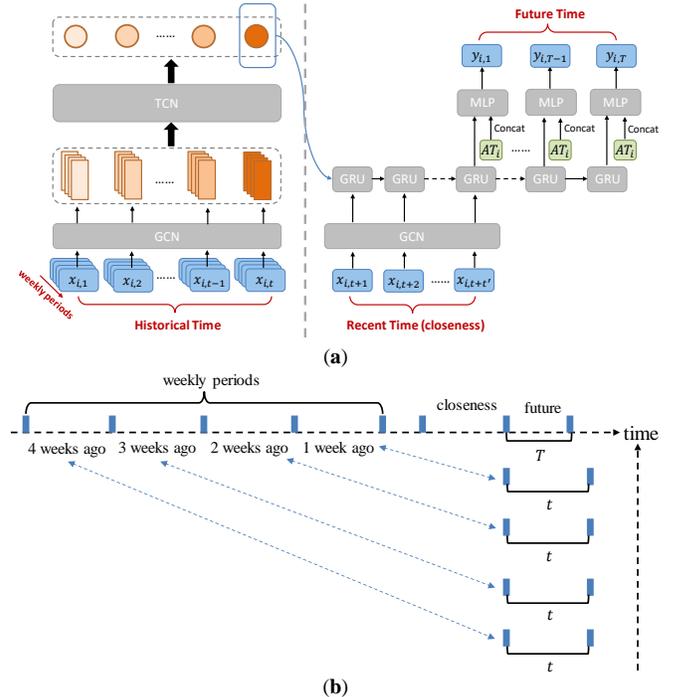

Fig. 2. (a) The proposed PSTN framework. (b) The generation of historical weekly periods.

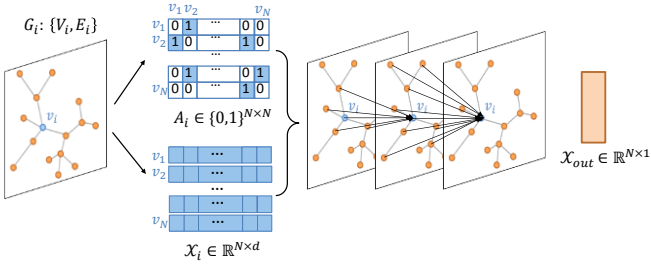

Fig. 3. The subgraph based on central road segment $v_i$ is $G_i$. With a two-layer GCN, the output for each sub-graph at a specific time is a vector with dimension $N$.

The number of layers in GCN indicates the farthest distance that node features can be gathered. For example, with a 1-layer GCN, each node can only obtain the feature information from the neighboring nodes. It happens independently at the same time for all the nodes. When stacking another layer on top of the first layer, the gathering process is repeated. However, in this time, the nodes already have information about their own neighbors (from the previous step). Thus, the second layer is able to gather information with a wider range related with central road $v_i$. Nevertheless, a deep architecture with many convolutional layers stacked may hurt the model performance [28]. Usually, a 2- or 3-layer GCN can obtain notable results in various tasks.

Although the spatial dependency for different road segments is dynamic, we fix the size of the subgraph to provide as many spatial relations as possible and avoid going through the entire topology graph. The algorithm for deciding a subgraph $G_i$ with fixed $N$ road segments is illustrated in Algorithm 1. The function $A(v_i, r_j)$ denotes the adjacency relation between road augment $v_i$ and $r_j$. As stated in Section 3.1, the result of $A(v_i, r_j)$ is 1 when $v_i$ and $r_j$ are adjacent, and 0 when they are not adjacent.

*2) Temporal Dependency Modeling*

Acquiring the temporal dependency is another key issue in the traffic prediction. There are two parts for temporal dependency modeling based on different types of the sequence. One is the historical weekly period. As presented in the left part of Fig. 2 (a), the historical information of length $t$ passes through GCN firstly to obtain the input matrix of TCN:

$$\mathcal{X}_i = \begin{bmatrix} \begin{bmatrix} x_{i,1,1}^1 & \cdots & x_{i,1,t}^1 \\ \vdots & \ddots & \vdots \\ x_{i,N,1}^1 & \cdots & x_{i,N,t}^1 \end{bmatrix}, \begin{bmatrix} x_{i,1,1}^2 & \cdots & x_{i,1,t}^2 \\ \vdots & \ddots & \vdots \\ x_{i,N,1}^2 & \cdots & x_{i,N,t}^1 \end{bmatrix}, \ldots, \\ \begin{bmatrix} x_{i,1,1}^w & \cdots & x_{i,1,t}^w \\ \vdots & \ddots & \vdots \\ x_{i,N,1}^w & \cdots & x_{i,N,t}^w \end{bmatrix} \end{bmatrix}, \quad (3)$$

where $N$ is the number of nodes in the subgraph $v_i$, and $w$ is the number of historical weeks folded according to the time. The selection of historical time is based on the future time that needs to be predicted. For example, the target prediction $\{y_{i,1}, y_{i,2}, \ldots, y_{i,T}\}$ decide that the same time slot $(1,2, \ldots, T)$ from $w$ weeks ago to one week ago are included in the historical time. As illustrated before, the length of historical time should be longer than that of future time, which is $t \geq T$.

The working process of TCN is displayed in Fig. 4. Based on different input length $t$, the number of layers stacked in TCN could be different. Each layer follows the convolution operation as:

$$H_l = \sigma_l(W_l * X_{l-1} + b_l), \quad l = 1,2, \ldots, L, \quad (4)$$

where $*$ denotes the convolution, $X_{l-1}$ is the output from last layer, $W_l$ and $b_l$ are the learnable parameters in layer $l$ and $\sigma_l$ is an activation function to increase the non-linearity in the TCN model. There are two differences between the TCN and conventional convolution. One is that the padding is only applied on the left side of the input sequence, making sure that each element from the output sequence only relies on its history. Another difference is the dilated convolution in TCN, which makes it advantageous in processing the time sequence. As shown in Fig. 4, the dilation in TCN refers to the distance between the elements of the input sequence that are utilized to compute the same output value. We set the dilation $d$ of the $i$-th layer in TCN as $d = 2^{i-1}$. With multiple dilated convolutional layers stacked, the output sequence of TCN has a full history coverage of the input sequence. To avoid the invalid temporal dependency from zero-paddings as much as possible, only the last element from the output is fed into GRU, which is the next sequence processing model.

---

**Algorithm 1** Deciding a subgraph $G_i$

**Input:** A central road segment $v_i$, number of nodes $N$, topology graph $G$

**Output:** A subgraph $G_i$.

1. Initialize $V_i = \{v_i\}, E_i = \{\}$
3. **for** all road segments $v_i \in V_i$ **do**
4.     **for** all road segments $r_j \in G$ **do**
5.         $e_{ij} = A(v_i, r_j)$
6.         **if** $e_{ij} = 1$ **then**
7.             $V_i = V_i \cup r_j, E_i = E_i \cup e_{ij}, G_i = \{V_i, E_i\}$
8.         k = count_node($V_i$)
9.         **if** k $\geq N$ **then**
10.             **return** $V_i, G_i$
11.         **end if**
12.     **end if**
13.     **end for**
14. **end for**
15. go back to step 3

GRU model is applied to model the second type of sequence, the recent time sequence. It employed the gated mechanism to memorize as much long-term information as possible, providing a solution to the problems of the long-term memory and exploding the gradient. The architecture of the GRU model is presented in Fig. 5. Consider states at time $t$ as an example, there are two types of gates in GRU cells including the reset gate and update gate. The reset gate $r_t$ is used to decide whether the previous cell state $h_{t-1}$ is important or not while the update gate $z_t$ decides if the cell state $h_t$ should be updated with the candidate state $c_t$ or not. The calculation process of a cell in the GRU is shown in (5) and (6). The $x_{i,t} \in \mathbb{R}^{1 \times d}$ is the output from GCN at the $t$-th step; $W_t \in \mathbb{R}^{3m \times (d+m)}$ and $b_t \in \mathbb{R}^{3m \times 1}$ are parameters of affine transformation; $\sigma$ and $tanh$ refer to the sigmoid and tanh activation functions, and $\odot$ denotes the element-wise product.

$$\begin{bmatrix} z_t \\ r_t \\ c_t \end{bmatrix} = \begin{bmatrix} tanh \\ \sigma \\ \sigma \end{bmatrix} (W_t \begin{bmatrix} (x_{i,t})^T \\ h_{t-1} \end{bmatrix} + b_t) \quad (5)$$

$$h_t = (1 - z_t) \odot h_{t-1} + z_t \odot c_t, \quad (6)$$

Different from the historical time, the recent time is not folded into weekly periods. Nevertheless, they both pass through the GCN model to generate spatial representations of the road network at each timestamp. Then, these time-varying representations are fed into the TCN and GRU models, respectively, to capture the temporal dependencies.

*3) Auxiliary Information Modeling*

As displayed in Fig. 2, the auxiliary information, such as attributes of road segments in the topology network, are concatenated at the end of GRU to predict traffic conditions through MLP. In this way, the model's perception of the specific road information is enhanced, thereby improving the prediction performance of certain road segment. We use $AT_i$ to denote the attributes vector of a certain road segment $v_i$. Each prediction $h_t \in \mathbb{R}^{m \times 1}$ is simply concatenated with $AT_i \in \mathbb{R}^{d' \times 1}$ and together fed into MLP, which is composed of multiple feed-forward layers and each follow Equation (4) to finely project the features. Considering the goal of traffic condition predictions is to directly predict the categorical congestion level, we apply the softmax activation and cross-entropy loss at the end of MLP:

$$f(z)_i = \frac{e^{z_i}}{\sum_{j=1}^{C} e^{z_j}}, \quad (7)$$

$$Loss = -\omega \sum_{i=1}^{C} k_i \log(f(z)_i), \quad (8)$$

where $f(z)$ refer to the activation function, $C$ is the number of classes (3 in our experiments), $\omega$ is the manually set weight tensor for different classes, and $k_i$ is a ground-truth indicator for class $i$. Algorithm 2 outlines the training process of PSTN.

## IV. COMPUTATIONAL EXPERIMENTS

In this section, we conduct computational experiments on two public traffic datasets to demonstrate the effectiveness of PSTN. We first introduce the datasets and the metrics used for the model evaluation. Then, we provide descriptions on the training setup with the model structure, following which, the ablation study with quantitative evaluation is presented. Finally, we compare the experimental results with several baselines to further reveal the advantages of PSTN model.

### A. Dataset and Evaluation Metrics

Two different traffic data are used in our experiments. Both of them have the categorical congestion level as the ground-truth label.

- Dataset1: The first dataset is an urban areal dataset in Beijing named "MapBJ"[1]. It contains 349 locations with more than 5 million records collected from March 2016 to June 2016 (two and half months) with a sampling interval of 5 minutes. The time-varying traffic features in dataset1 include four fields, the road segment id, timestamp, traffic condition, and the

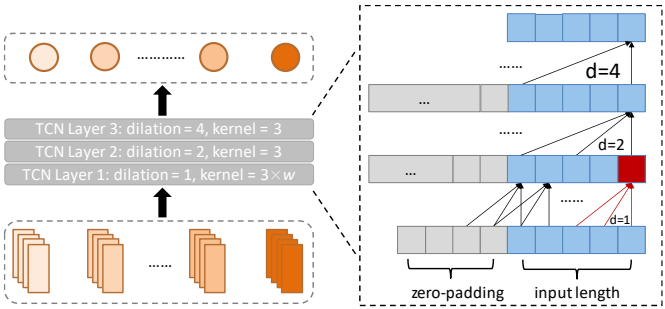

Fig. 4. The architecture of TCN for processing the historical weekly period.

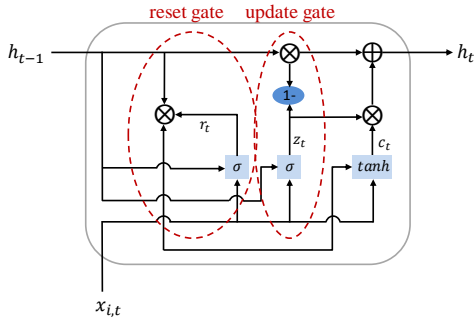

Fig. 5. The structure of a GRU cell.

[1] https://github.com/cxysteven/MapBJ

**Algorithm 2** Training process of PSTN.

**Input:** Ground-truth of congestion levels $C$, training set $X_{train}$, max number of epochs $E$.

**Output:** The learned model with parameters $\Theta$.

1. Initialize all learnable parameters $\Theta_0$.
2. **for** epoch $e = 1, ..., E$ **do**
3.     Sample a batch of training instances $X \in X_{train}$.
4.     $\mathcal{X} = GCN(X)$
5.     $Z = MLP(TCN(\mathcal{X}), GRU(\mathcal{X}))$
6.     update $\Theta$ by minimizing the objective (7) & (8) with $Z$
7.     **repeat** step 3-6 **until** $X_{train}$ is all trained.
15. **end for**

speed limit level. Specifically, the traffic condition owns 5 values described in a set {not released, unblocked, slow, congested, extremely congested}, which is next coded to {0, 1, 2, 3, 4}. There is also a topology network in dataset1 with road ids. We utilize the first two months data for training and the remaining half month for the validation and testing. Considering that the speed limit is an important reference for road segmentation, it is treated as one of the traffic features as well as the road attributes in our experiments.

- Dataset2: The second dataset is also a public dataset with more than 14 million records in Xi'an, provided by Didi Chuxing[2]. The time period of this dataset is from July 1$^{st}$ 2019 to July 30$^{th}$ 2019, taking two minutes as a sampling interval. Although the duration of the records in dataset2 is not as long as dataset1, the records themselves already contain the historical time and recent time. For example, with a specific road id and the future time, the traffic features at the recent 5 time slices and historical 5 time slices from 4 weeks ago are provided. The time-varying traffic features contain the speed, eta speed, the number of cars, and the traffic condition. Similarly, the traffic condition in dataset2 is categorized into 4 classes {unblocked, slow, congested, extremely congested}, corresponding to {1, 2, 3, 4}. The time-invariant traffic information including the road network and road attributes are also provided in dataset2. We use the data from July 1$^{st}$ to July 28$^{th}$ for training, the records on July 29$^{th}$ for validation and July 30$^{th}$ for testing.

Fig. 6 is the histogram of the two datasets. We merge the condition 3 with condition 4 as they both represent a congested traffic status only with different degrees. As displayed in Fig. 6, there is a severe imbalance between different traffic conditions. The condition 1 representing the unblocked traffic occupies most of the records, followed by condition 2, and the condition 3, which really represents the congestion, only accounts for a small proportion. Based on the class imbalance, we adopt the weighted cross-entropy when calculating the loss to ensure that the loss of majority class will not overwhelm the minority class. From the perspective of the timestamp, dataset1 appears to be more regular while dataset2 has more scattered future time.

For all experiments, we exploit the following common metrics, the accuracy, precision, recall, and F1-score. They are defined as follows:

$$accuracy = \frac{TP + TN}{TP + TN + FP + FN}, \quad (9)$$

$$precision = \frac{TP}{TP + FP}, \quad (10)$$

$$recall = \frac{TP}{TP + FN}, \quad (11)$$

$$F1 = \frac{2 \times precision \times recall}{precision + recall}, \quad (12)$$

where $TP$ and $TN$ are the correctly predicted positive and negative values, respectively; $FP$ and $FN$ are the wrongly predicted values. These 4 types of predictions make up a confusion matrix, which will be shown in the next content. Accuracy is simply a ratio of correct predictions to the total predictions and it is the most intuitive performance measurement. However, accuracy is a great measure only when the classes are balanced, i.e., the values of $FP$ and $FN$ are almost the same. In our experiments, the three categories of traffic condition are imbalanced, in which the case precision and recall are introduced to better interpret the prediction results. Precision quantifies the ratio of correctly predicted positives to the total predicted positives. A high precision value relates to the low false positive rate. Recall is the ratio of correctly predicted positives to the all actual positives. A high recall value indicates that the model can recognize most of positive samples without mistakenly recognizing negative examples as positive. The baseline for recall is 0.5. Finally, the F1-score is defined as the harmonic mean of the precision and recall.

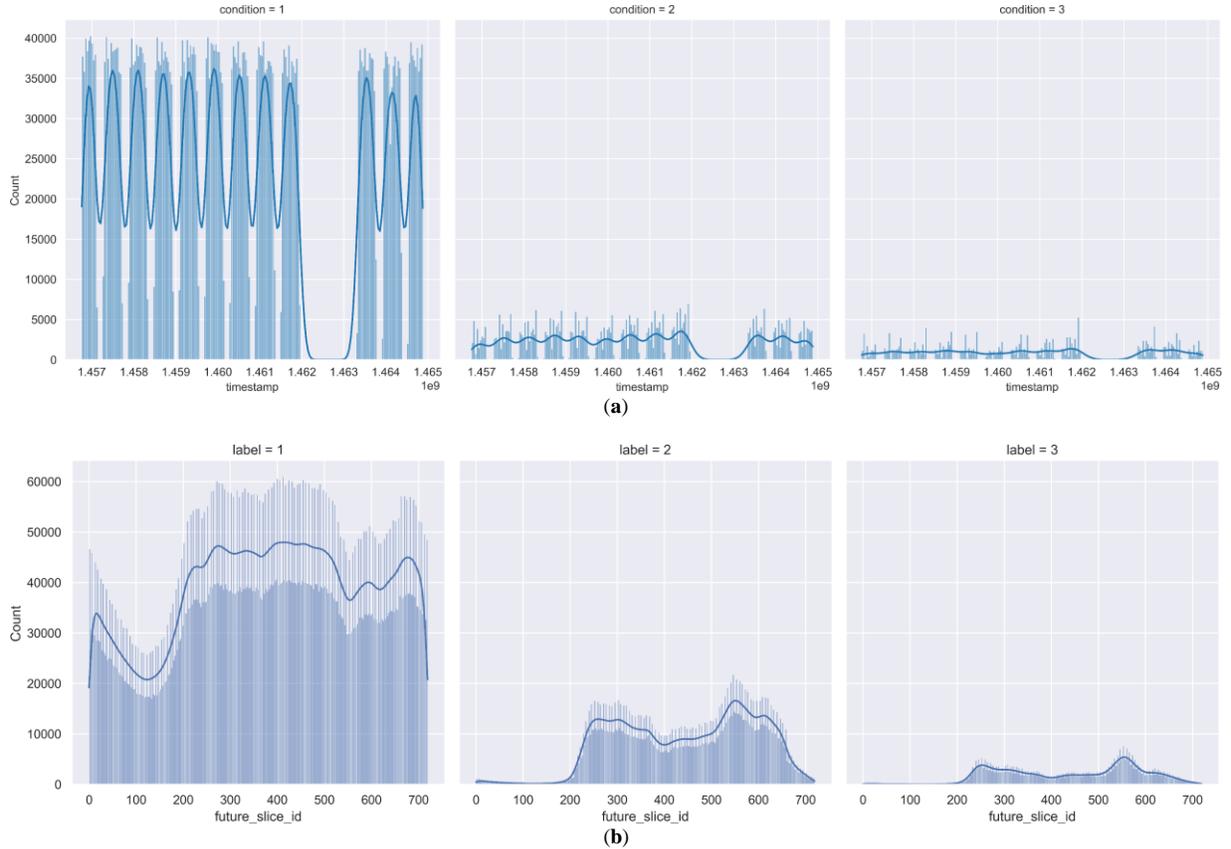

Fig. 6. (**a**) Histogram of traffic conditions from dataset1. (**b**) Histogram of traffic conditions from dataset2.

---

[2] https://gaia.didichuxing.com

*B. Experimental Settings*

The hyperparameters in different modules of the PSTN model are also different. In GCN module, we select the size of subgraph $G_i$, which is also the number of nodes in $G_i$, from the candidate set $\{50, 100, 150, 200\}$ and analyze the change of prediction performance. In TCN modules, the lengths of historical time $t$ in two datasets are different. Dataset2 has a fixed historical length of 5 time slices while the $t$ of dataset1 is set as $t \in \{6, 9, 12\}$. Similarly, in GRU module, the length of recent time $t'$ in dataset2 is fixed as 5 while in dataset1 it is optional and we set $t'$ as $t' \in \{6, 9, 12\}$, just like the historical length. The prediction horizon $T$ is also explorable in our experiments. Since the future time in dataset2 has been assigned for each instance, we can only change the prediction horizon in dataset1. Considering the time interval in dataset1 is 5 minutes, we set $T \in \{1, 3, 6, 9, 12\}$ (corresponding to 5, 15, 30, 45, 60 minutes after) to evaluate the robustness of the PSTN method. The hidden dimension in GRU is fixed to $m = 64$. Additionally, the number of hidden units in MLP layers changes in the order of $[256, 512, 1024, 512, 256]$ and finally output a 3-dimension prediction.

In the experiment, we manually adjust and set the learning rate to 0.001, the batch size to 1024, and the training epoch to 300. The model is trained using the AdamW optimizer. The deep networks are implemented in Pytorch. Experiments are conducted using three NVIDIA GeForce RTX 2080Ti GPUs.

Taking 5-min prediction ($T = 1$) as an example, Fig. 7 shows the F1-score of the validation set from dataset1 with different $t$, $t'$ and $G_i$. As the subgraph $G_i$ continues to expand, the optimal length combination of $t$ and $t'$ is also changing. For instance, when only 50 nodes are contained in $G_i$, $t = 6$ and $t' = 6$ are able to yield the highest F1-score while, when there are 200 nodes in $G_i$, the combination of $t = 6$ and $t' = 6$ can only yield the lowest F1-score. In general, the larger the subgraph, the longer the time length is required to achieve a comparable prediction performance. By comparing the F1-score under different length combinations, we decide the historical length and recent length as $t = 9$ and $t' = 12$, respectively, to conduct the further computational experiments on dataset1. Fig. 8 summarizes the validation F1-score of two datasets with different size of $G_i$. The optimal size of $G_i$ for dataset1 is 100 nodes with $t = 9$ and $t' = 12$, and the optimal size for dataset2 is 150 nodes with $t = 5$ and $t' = 5$. Through the preliminary experiments, the hyperparameters are carefully designed with consideration of the balance between the model complexity and computational load of the overall system.

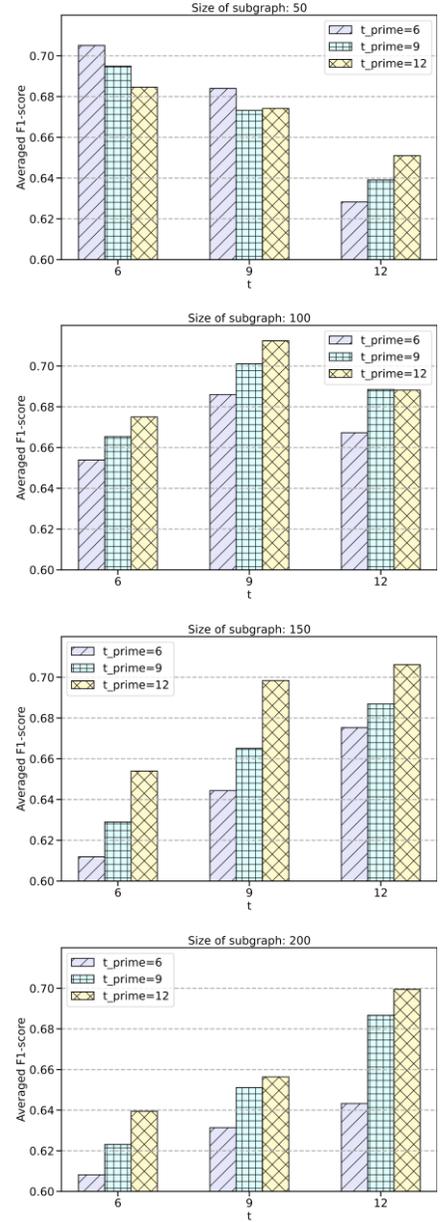

Fig. 7. Prediction performance for dataset1 with different $t$, $t'$ and $G_i$.

*C. Ablation Studies*

To demonstrate the utility of the different components of the proposed methodological framework, we performed ablation studies by systematically removing each component in turn. As stated in Section 3.2, we divide the whole framework into three components, the combination of TCN and GCN that takes historical weekly periods as input, the combination of GRU and GCN that takes recent time as input, and the MLP that takes additional road attributes as input. Specifically, the GRU module taking the recent traffic features after GCN as input is treated as a baseline. The experimental setting is divided into adding only weekly periods (WP) information, adding only road attributes (AT) information, as well as adding both WP and AT. Our full model includes both WP and AT. Testing results of ablation studies are summarized in Table 1, which illustrates that the historical weekly periods and road attributes can play a role in the task of traffic condition prediction. From the perspective of simply adding AT information, the F1-score is increased by approximately 0.03 for dataset2 compared with the baseline. From the perspective of simply adding WP information, the F1-score in dataset1 is 0.09 higher than the baseline. After adding both AT and WP, the F1-score is increased by approximately 0.1 and 0.07 for dataset1 and dataset2, respectively. When considering the type of information to introduce, the model enhanced with WP information performs better than the model with AT information, indicating that the impact of historical time on traffic condition is greater than that of the road attributes. Moreover, with both AT and WP information, the model yields higher accuracy than the model that adds only a single type of information. The prediction error is lower up to 0.08, indicating the complementarity of the two components. In general, enhancing the model with WP and AT can facilitate the prediction task.

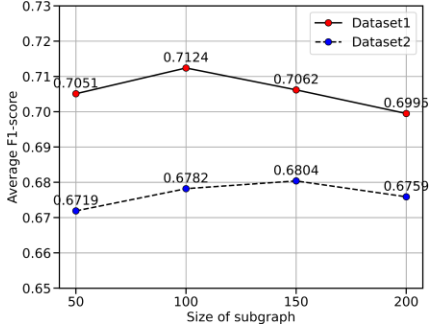

Fig. 8. The average F1-score of dataset1 and dataset2 based on different size of $G_i$.

The confusion matrixes on two test sets are presented in Fig. 9. Due to a high proportion of samples, the condition 1 (unblocked) owns the greatest accuracy values on both datasets (0.8461 and 0.7685, respectively) while the condition 3 (congested) has the lowest prediction accuracies of 0.0171 and 0.0189, respectively. Table 2 is the metrics summary of the confusion matrixes. Although the accuracies of the two test sets are quite different, considering the severe imbalance between three conditions, we are mainly concerned with the performance on average F1-score, which can better balance the precision and recall on different categories. The average recall values over three categories on both datasets are greater than 0.5, the baseline, which indicates that the model performs well when $T = 1$. Additionally, the average precision values around 0.7 indicate relatively low false positive rates in both datasets.

### D. Comparison Results and Analysis

In this section, we verify the effectiveness of our proposal by comparing it with several state-of-the-art methods for predicting traffic congestion levels within various prediction horizons. To ensure a fair comparison, a common test set, the same input information and measures of performance are used. These state-of-the-art methods are listed as follows:

- GRU [29] is a classic variant of RNN that shows superior capability for the time series prediction with a long temporal dependency. A GRU cell is composed of an update gate and a reset gate, which is simpler and requires less training time than LSTM model. In this experiment, the historical weekly periods are flattened and concatenated with the recent time to pass through the GRU cells.

**Table 1.** Ablation studies on two datasets.

| Model | Precision | Recall | F1-score |
|---|---|---|---|
| *Dataset1* | | | |
| Baseline | 0.6173 | 0.6151 | 0.6162 |
| Baseline + AT | 0.6254 | 0.6304 | 0.6279 |
| Baseline + WP | **0.7073** | 0.7031 | 0.7051 |
| Baseline + AT + WP | 0.7012 | **0.7208** | **0.7107** |
| *Dataset2* | | | |
| Baseline | 0.6031 | 0.6073 | 0.6052 |
| Baseline + AT | 0.6370 | 0.6303 | 0.6336 |
| Baseline + WP | 0.6491 | 0.6429 | 0.6458 |
| Baseline + AT + WP | **0.6762** | **0.6758** | **0.6760** |

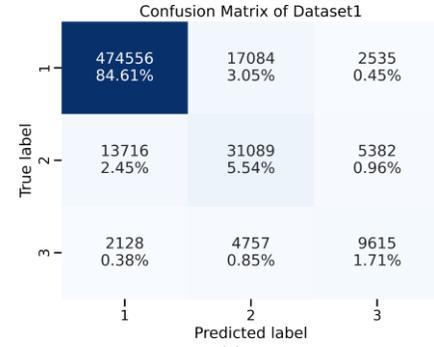

(a)

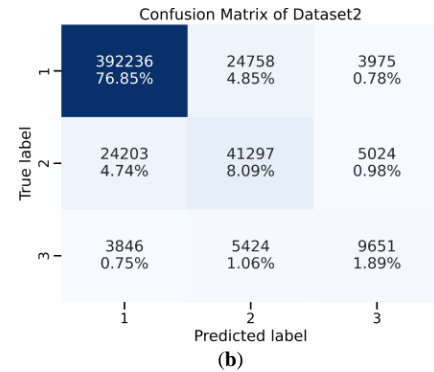

(b)

Fig. 9. (a) Confusion matrix of dataset1. The hyperparameters are set as $t = 9$ and $t' = 12$, and size of $G_i = 100$. (b) Confusion matrix of dataset1 and dataset2. The hyperparameters are set as $t = 5$ and $t' = 5$, and size of $G_i = 150$.

**Table 2.** Results summary of the confusion matrixes.

|  | Dataset1 | | | Dataset2 | | |
| --- | --- | --- | --- | --- | --- | --- |
| | Precision | Recall | F1-score | Precision | Recall | F1-score |
| 1 | 0.9677 | 0.9603 | 0.9640 | 0.9333 | 0.9317 | 0.9325 |
| 2 | 0.5874 | 0.6195 | 0.6030 | 0.5778 | 0.5856 | 0.5817 |
| 3 | 0.5484 | 0.5827 | 0.5827 | 0.5175 | 0.5101 | 0.5138 |
| Accuracy | / | / | 0.9186 | / | / | 0.8683 |
| Average | 0.7012 | 0.7208 | **0.7107** | 0.6762 | 0.6758 | **0.6760** |

**Table 3.** Comparison with benchmarks on the testing data from dataset1. The training schedules are all set to 300 epochs to make sure good baselines.

| Model | Dataset1 | | | | | Dataset2 |
| --- | --- | --- | --- | --- | --- | --- |
| | 5 min | 15 min | 30 min | 45 min | 60 min | / |
| GRU | 0.6141 | 0.6120 | 0.6082 | 0.5931 | 0.5927 | 0.5873 |
| T-GCN | 0.6312 | 0.6296 | 0.6295 | 0.6130 | 0.6124 | 0.6230 |
| DCRNN | 0.6418 | 0.6323 | 0.6300 | 0.6252 | 0.6250 | 0.6482 |
| ST-3DNet | 0.6650 | 0.6594 | 0.6488 | 0.6379 | 0.6326 | 0.6595 |
| PSTN | **0.7107** | **0.6954** | **0.6813** | **0.6745** | **0.6737** | **0.6760** |

- T-GCN [27] simply combines the GCN and the GRU to do the traffic forecasting. It is similar with the baseline structure we applied in ablation studies. Nevertheless, to ensure a fair comparison in this section, the historical information as well as the road attributes are employed in the training process.
- DCRNN [26] models the traffic information as a diffusion process. It captures the spatial dependency using bidirectional random walks on the topology graph, and the temporal dependency using the encoder-decoder architecture.
- ST-3DNet [4] is based on 3D convolution and recalibration block to model the periodic traffic data and space sorrelation, respectively. Different from our proposal, the weekly periods and closeness are fused together through 3D convolution for short-term traffic congestion prediction.

Except for GRU, the pure temporal model, other methods take the usage of spatial dependency in different ways and result in spatial-temporal frameworks. The hyperparameters in the baselines above are optimized by a grid search. For instance, we vary the hidden dimension in GRU based on a candidate set $[16, 32, 64, 100, 128]$, and find that it obtains the best performance when the hidden dimension is set to 64. Considering these state-of-the-art methods were proposed based on various traffic prediction tasks, a common MLP architecture is connected at the end of models to output the categorical congestion levels. The training schedules are all set to 300 epochs to make sure good baselines. To compare the PSTN with the baselines on dataset1, we not only evaluate the performance when prediction horizon equals 5 min, but also measure it under even longer prediction horizons. As the prediction horizon becomes longer, the optimal length of historical time may not be 9. A basic requirement for the length of historical time is $t \geq T$, which means that the target future time must be included in the historical periods. We report the comparison results in terms of F1-score on testing data from two datasets in Table 3, where we see that our proposed method is competitive with state-of-the-art methods. On the whole test sets, the results for T-GCN and DCRNN are similar. One potential reason is that the flattened historical information is concatenated before recent time information in T-GCN and DCRNN, only the recent information can be remembered. The other reason is the flattened weekly periods might be harmful for learning the periodic temporal dependency. We can see that ST-3DNet performs better than T-GCN and DCRNN, indicating the effectiveness of considering the long-term and short-term information separately. Compared with the best baseline results, PSTN achieves approximately 0.02 higher performance in terms of F1-score in dataset2. On dataset1, the improvements are even more significant. In terms of various prediction horizons, we get approximately 0.04 improvements for each horizon compared with the best baseline results. These results demonstrate the effectiveness of our proposed PSTN model.

## V. CONCLUSION

In this work, we proposed a novel deep learning based spatial-temporal network named PSTN for predicting traffic condition. We utilized GCN to capture the spatial dependencies of nodes in the subgraph network. At the same time, TCN and GRU were employed to capture both historical and local information of traffic sequences in the temporal dimension. Specifically, we folded the historical sequence data and reconstructed it into a three-dimensional matrix considering the characteristics of urban traffic congestion data. Furthermore, the auxiliary information of road attributes was concatenated to enhance the prediction performance. Our proposal was evaluated on two different datasets in the traffic domain. The experimental results showed that our model is capable to perform robustly on various prediction horizons and outperformed the state-of-the-art baselines significantly for the traffic condition prediction.

ACKNOWLEDGMENT

Thanks for the support of data source from Didi Chuxing GAIA Initiative.